# Excited state dynamics of liquid water:

# Insight from the dissociation reaction following two-photon excitation

Christopher G. Elles, Ilya A. Shkrob, and Robert A. Crowell[a]

*Chemistry Division, Argonne National Laboratory, Argonne, Illinois 60439*

Stephen E. Bradforth[b]

*Department of Chemistry, University of Southern California, Los Angeles, California 90089*

We use transient absorption spectroscopy to monitor the ionization and dissociation products following two-photon excitation of pure liquid water. The two decay mechanisms occur with similar yield for an excitation energy of 9.3 eV, whereas the major channel at 8.3 eV is dissociation. The geminate recombination kinetics of the H and OH fragments, which can be followed in the transient absorption probed at 267 nm, provide a window on the dissociation dynamics at the lower excitation energy. Modeling the OH geminate recombination indicates that the dissociating H atoms have enough kinetic energy to escape the solvent cage and one or two additional solvent shells. The average initial separation of H and OH fragments is 0.7±0.2 nm. Our observation suggests that the hydrogen bonding environment does not prevent direct dissociation of an O-H bond in the excited state. We discuss the implications of our measurement for the excited state dynamics of liquid water and explore the role of those dynamics in the ionization mechanism at low excitation energies.

[a] Electronic mail: rob_crowell@anl.gov
[b] Electronic mail: stephen.bradforth@usc.edu

## I. INTRODUCTION

The excited state dynamics of liquid water play an important role in a wide range of applications. A prominent example is the irradiation of aqueous systems, where ionized and electronically excited water molecules decompose into reactive species that are responsible for much of the subsequent chemistry.[1] Although much is known about the kinetics of the transient species, the initial dynamics that govern their formation have not been fully characterized. In this paper we examine the photodissociation of liquid water and consider the role of excited state dynamics in the liquid-phase ionization channel.

The isolated water molecule is a useful reference for studying the liquid. The maximum of the first absorption band is near 7.5 eV in the gas phase, corresponding to a transition that promotes an electron from the highest occupied non-bonding valence orbital ($1b_1$) to an unoccupied anti-bonding orbital ($4a_1/3s$) with significant Rydberg character. This first electronic excited state, which is well below the 12.6 eV ionization potential of the isolated molecule, is strongly repulsive and promptly dissociates to H and OH with a large fraction of the available energy going into relative translation of the fragments.[2,3] The second excited state corresponds to a transition from the next highest valence orbital ($4a_1/3s \leftarrow 3a_1$) and also dissociates along an O-H bond. Much less is known about the excited state dynamics of liquid water, although a shift of the first absorption band by about 0.7 eV to higher energy indicates that the liquid environment has a strong influence on the electronic structure.[4]

A particularly intriguing aspect of the dynamics of liquid water is the role that hydrogen bonding plays. Hydrogen bonding gives water many of its unusual properties and recent work using multidimensional vibrational spectroscopy, among other techniques, has uncovered many interesting aspects of how hydrogen bonding affects the ground state dynamics.[5-7] There is far less work addressing the role of cooperative effects on the excited state dynamics of the liquid.[8,9] Furthermore, because it is difficult to accurately include cooperative effects in electronic structure calculations, only a limited number of theoretical treatments concerning the electronically excited states of condensed-phase water are available.[10-15] Even fewer calculations explore excited state potential energy surfaces outside of the Franck-Condon region, and these are generally limited to the dimer and other small water clusters.[16-21] Experimental measurements of the dynamics in the bulk liquid provide important benchmarks for comparison with theoretical and computational results.

Previous investigations of the dynamics following electronic excitation of liquid water generally focus on the mechanism of ionization, rather than dissociation.[22-26] However, early studies irradiating water with vacuum ultraviolet light indicate that both processes play a role in the liquid.[27-30]

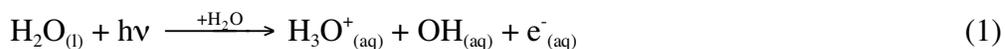
$$H_2O_{(l)} + h\nu \xrightarrow{+H_2O} H_3O^+_{(aq)} + OH_{(aq)} + e^-_{(aq)} \quad (1)$$

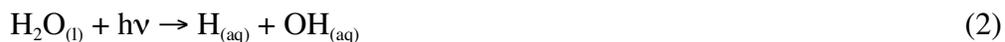
$$H_2O_{(l)} + h\nu \rightarrow H_{(aq)} + OH_{(aq)} \quad (2)$$

Unlike in the gas phase, ionization occurs down to the onset of optical absorption near 6.5 eV for both one- and two-photon excitation of liquid water.[30,31] The energy for *vertical* ionization of the liquid is about 11 eV, and the discrepancy implies that nuclear motion must play a role in the ionization mechanism at low energies.[26] The precise relationship of dissociation and ionization, whether they are mutually exclusive pathways or slightly different outcomes of a similar process,[32] is an unresolved distinction that lies at the heart of understanding the ionization mechanism in this energy regime. Our experiments continue to unravel these details by explicitly examining the excited state dynamics of liquid water.

The work described in this paper uses transient absorption spectroscopy to monitor the products of ionization and dissociation. The dissociation channel plays the more important role for an excitation energy of 8.3 eV, while dissociation and ionization occur with roughly equal probability for 9.3 eV excitation. Our result agrees with previous observations that the ionization yield increases rapidly across this range of energies,[31,32] presumably with a corresponding decrease in the dissociation yield.[33] At the lower excitation energy, where the dissociation channel dominates, we monitor the geminate recombination kinetics of the dissociation products and determine that the average initial separation of H and OH fragments is about 0.7 nm, similar to the value from recent molecular dynamics simulations of dissociating $H_2O$ in ice.[34,35] The relatively large separation of the dissociation products suggests that H atoms are formed with enough kinetic energy to escape the solvent cage and one or two additional solvent shells.

## II. EXPERIMENT

Time-resolved transient absorption measurements monitor the evolution of the ionization and dissociation products following two-photon excitation of liquid water. We probe the transient products at two wavelengths, 650 and 267 nm. The solvated electron is the only species that absorbs visible light, and therefore the transient signal that we measure at 650 nm reflects purely the geminate recombination kinetics following ionization. In contrast, the signal at 267 nm potentially includes contributions from the kinetics following dissociation as well as ionization, because both electrons and OH radicals absorb at that wavelength. Hydrogen atoms and hydronium ions do not absorb light at either wavelength.

In addition to the measurements for pure water, we observe the transient change in absorption for a 2M solution of perchloric acid, where protons from the acid rapidly react with solvated electrons from the ionization of water. By eliminating solvated electrons, the measurements in acid solution reveal the relative contribution of OH radicals to the ultraviolet absorption signal. Comparing the transient absorption for pure water and a 2M solution of NaClO$_4$ confirms that irradiation of perchlorate ions does not contribute to the signal because the transient absorption is the same in the two solutions. Perchlorate ions are the only likely source of additional electrons in the acid and salt solutions. We use deionized water with greater than 18 MΩ/cm resistance for the pure water samples, and obtain 2M acid solutions by diluting reagent grade 70% HClO$_4$ (Sigma-Aldrich).

The ~100 fs excitation and probe pulses come from frequency conversion of the 800 nm light from an amplified Ti:sapphire laser consisting of an oscillator (Spectra Physics, Tsunami) and two consecutive multi-pass amplifiers. The laser system produces 1.6 mJ pulses with a 1 kHz repetition rate, and we use up to 90% of the 800 nm light to generate excitation pulses. Frequency quadrupling the signal output of an optical parametric amplifier (Spectra Physics, OPA 800C) gives excitation pulses at 300 nm (4.13 eV), whereas frequency tripling the Ti:sapphire fundamental gives pulses at 267 nm (4.65 eV), in each case providing up to 3 μJ per pulse. A small fraction of the remaining 800 nm light passes onto a computer-controlled delay stage before we use it to produce probe pulses. Visible probe pulses come from generating white-light continuum in a water cell and passing it through an interference filter with a center wavelength of 650 nm, ultraviolet probe pulses come from frequency tripling the 800 nm light.

Two lenses separately focus the excitation and probe beams into a 100 μm thick gear-pumped liquid jet, where they intersect at a small angle. Typical beam diameters at the sample are 30-100 μm for the excitation beams and 15-30 μm for the probe beams. Smaller beam diameters for 8.3 eV excitation are necessary to compensate for the smaller two-photon absorption cross-section compared with 9.3 eV excitation. We monitor the intensity of the probe before and after the sample, to account for fluctuations of the laser, and a chopper wheel blocks every other excitation pulse in order to measure the change in optical density as a function of the delay between excitation and probe pulses. The transient absorption signal at each probe wavelength changes quadratically with the intensity of the excitation pulse, confirming that the products come from two-photon excitation.

## III. RESULTS AND ANALYSIS

**A. Two-photon excitation at 9.3 eV**

A comparison of the transient absorption signals at 650 and 267 nm provides a quantitative measure of the relative product yields from ionization and dissociation. Figure 1 shows the change in absorption at the two probe wavelengths following biphotonic excitation of liquid water at 9.3 eV. The traces in the inset are normalized over the range from 5 to 20 ps to show that the signal decays by the same amount for each probe wavelength (see below). For delay times longer than 5 ps the absorption signal is 9.8±1.7 times stronger at 650 nm than it is at 267 nm, where the reported absorption ratio and uncertainty are the average value and standard deviation from a set of several independent measurements. The uncertainty primarily reflects variations in the spatial overlap of the probe and excitation beams in each measurement. Slightly different overlap for the two probe beams adversely affects the relative absorption measurement by causing the beams to sample different product concentration profiles.

The relative absorption strength at each probe wavelength, together with the absorption coefficients in Table I,[36,37] indicate that two-photon excitation of water at 9.3 eV produces more OH radicals than solvated electrons. For example, if ionization were the only channel the ratio of OH radicals and solvated electrons would be nearly equal and the absorption would be about 15 times stronger at 650 nm than 267 nm. From our measurement we calculate that the ratio of OH radicals to electrons is [OH]/[e⁻] = 2.3±0.7. Because ionization produces equal concentrations of solvated electrons and OH radicals, a large excess of the latter suggests that dissociation must also play a role. In fact, a slight excess of OH radicals over electrons is expected even for pure ionization because a small fraction (~20%) of recombining electrons react with the geminate $H_3O^+$ ion rather than the OH radical,[38,39] but the effect of the competing recombination reactions following ionization is too small to account for the large excess of OH radicals that we observe. Thomsen et al.[39] also measure the transient absorption following two-photon excitation at 9.3 eV and they obtain [OH]/[e⁻] = 1.55. Although their value is slightly smaller than ours, the two results are in reasonable agreement given the uncertainty of both measurements.

An alternate method of determining the relative contribution to the absorption signal from each of the products is to compare the transient absorption in pure water and in an acid solution.[40] The top panel of Fig. 2 compares the electron signal at 650 nm for pure water (closed circles) and for a 2M solution of $HClO_4$ (open circles). Ionization of pure water results in the usual geminate recombination kinetics, but the electron signal exponentially decays to the baseline within about 100 ps in the acid solution due to bimolecular reaction with excess protons from the strong acid ($k_3 = 1.2\times10^{10}$ $M^{-1}s^{-1}$ for a 2M solution of $HClO_4$).[41]

$$e^-_{(aq)} + H_3O^+_{(aq)} \rightarrow H_{(aq)} + H_2O_{(l)} \quad (3)$$

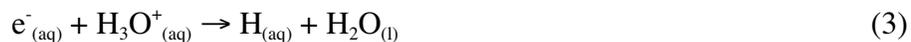

The data in the lower panel of the figure are the corresponding transient absorption traces at a probe wavelength of 267 nm. The ultraviolet absorption at long delay times is about 45% weaker for the acid solution relative to pure water, indicating that electrons account for about 45% of the transient signal. Because $H_3O^+$ ions and H atoms do not absorb at either 650 or 267 nm, the primary difference between the traces in pure water and in the acid solution is the additional decay of electrons in the latter.[42] Based on the absorption cross-sections of the products (Table I), the 45% contribution of electrons to the 267 nm signal gives [OH]/[e$^-$] = 1.7±0.9, in good agreement with the above value from the transient absorption data for pure water.

**B. Two-photon excitation at 8.3 eV**

Figure 3 shows the transient change in absorption at the two probe wavelengths following biphotonic excitation of pure water at 8.3 eV. Contrasting with the higher excitation energy, the decay is noticeably different for the traces in the inset, which are normalized over the range from 5 to 20 ps. The ratio of OH radicals to solvated electrons for delays longer than 100 ps is at least [OH]/[e$^-$] = 3.3±1.0, indicating that a larger fraction of excited molecules dissociate at 8.3 eV than at 9.3 eV. The uncertainty is larger in this case than at the higher excitation energy because we focus the excitation beam to a smaller diameter at the sample in order to compensate for the lower laser intensity and two-photon absorption cross-section of water at 8.3 eV. The overlap of the excitation and probe beams in the sample is more sensitive for a smaller excitation beam, which increases the uncertainty of the absorption measurement at the different probe wavelengths. The acid quenching experiment, on the other hand, is independent of the relative overlap of excitation and probe beams at different wavelengths because that measurement compares the transient data for pure water and acid solutions at a single wavelength.

The transient absorption data for 8.3 eV excitation of pure water and an acid solution are shown in Fig. 4. In contrast with the result at 9.3 eV, the 267 nm absorption signal for long delay times is the same in both water and 2M HClO$_4$. The similarity of the traces indicates that electrons do not make a significant contribution to the transient ultraviolet absorption at this excitation energy. Considering the signal-to-noise ratio in our experiment, we estimate that electrons contribute less than 15% of the 267 nm absorption signal following 8.3 eV excitation. Although the resulting ratio [OH]/[e⁻] > 8 is somewhat higher than we obtain from the relative absorption measurement in Fig. 3, we believe that the acid quenching experiment gives the more accurate estimate in light of the relatively large uncertainty in comparing the absorption strength at two probe wavelengths. Because ionization produces nearly equal concentrations of OH radicals and solvated electrons, the <15% electron contribution implies that OH radicals from the dissociation channel are responsible for more than 75% of the signal. We conclude that OH radicals from dissociation are the dominant absorbing species probed at 267 nm following 8.3 eV excitation, and that the dissociation yield is several times larger than the ionization yield at this energy. The different decay of the signal for each probe wavelength in the inset of Fig. 3 is a result of probing the products from different reactions; the solvated electron signal in the visible reveals the geminate recombination kinetics for the minor ionization channel, while recombining OH radicals formed via the dissociation channel dominate the signal in the ultraviolet.

## C. Models of geminate recombination

*Ionization*

Solvated electrons are the only species that absorb light at 650 nm, therefore the decay of the absorption signal at that wavelength reflects the recombination of electrons with their geminate partners following two-photon ionization of water.

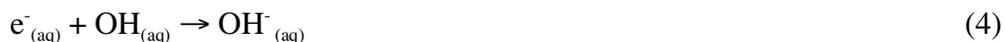

$$e^-_{(aq)} + OH_{(aq)} \rightarrow OH^-_{(aq)} \qquad (4)$$

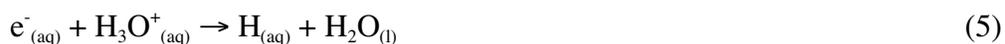

$$e^-_{(aq)} + H_3O^+_{(aq)} \rightarrow H_{(aq)} + H_2O_{(l)} \qquad (5)$$

We fit the electron signal at this probe wavelength using the independent pairs model, which assumes that the competing recombination reactions (4) and (5) proceed independently.[38,43] The independent pairs model accurately reproduces the signal decay using a single variable fit parameter, $<r_0>$, which is the average initial separation of solvated electrons and their ionization counterparts in thermal equilibrium with the solvent.[26] We assume a Gaussian distribution of electron ejection lengths, centered at the origin of the ionization site, with the value of $<r_0>$ related to the width of the distribution. Larger initial separations lead to less recombination because the geminate species are less likely to diffusively encounter each other. Fits to the 650 nm electron decay data using the independent pairs model (solid lines in Figs. 1 and 3) give average ejection lengths of $<r_0>$ = 1.0±0.2 and 1.4±0.2 nm for 8.3 and 9.3 eV excitation, respectively. These values are in excellent agreement with our previous work studying in detail the variation of the ejection length with excitation energy in the range from 7.8 to 12.4 eV.[25,26]

*Dissociation*

The absorption signal at 267 nm is sensitive to both solvated electrons and OH radicals. For example, the electron quenching experiment for 9.3 eV excitation suggests that about 45% of the ultraviolet absorption is from electrons, with the remaining signal from OH radicals formed by ionization (~32%) and dissociation (~23%). The decay of the absorption in pure water reflects primarily the kinetics of the recombining ionization products, but also includes a contribution from the decay of OH radicals following dissociation. We cannot extract any information about the dissociation channel from the 9.3 eV data because of the large contribution from ionization. At the lower excitation energy, however, the absence of an appreciable electron signal indicates a much smaller contribution of the ionization products to the 267 nm absorption. Instead, OH radicals from the dissociation of water are largely responsible for the kinetics, and the transient ultraviolet absorption provides a window on the dissociation reaction following 8.3 eV excitation. Dissimilar kinetics in the 267 and 650 nm absorption traces in the inset of Fig. 3 confirm that different recombination reactions contribute to the signal at each probe wavelength.

Figure 5 shows the transient decay of the 267 nm absorption signal following 8.3 eV excitation. We ignore the small contribution to the signal from the ionization channel and fit the data with a simple model for diffusion-limited recombination of the dissociation products.

$$H_{(aq)} + OH_{(aq)} \rightarrow H_2O_{(l)} \qquad (6)$$

The recombination model gives a time-dependent expression for the survival probability of OH radicals,

$$[OH](t) = [OH](0) \cdot \left[1 - \left(\frac{R}{r_{H-OH}}\right) \cdot \mathrm{erfc}\left(\frac{r_{H-OH} - R}{\sqrt{4Dt}}\right)\right] \qquad (7)$$

where $r_{H-OH}$ is the initial separation of a pair of fragments, $D$ is the sum of their diffusion rates, and $R$ is the reaction radius, within which the species recombine with unit yield.[44] The steady-state reaction rate constant ($k_6 = 4\pi N_A R D = 2.0 \times 10^{10}$ M$^{-1}$s$^{-1}$) and joint diffusion constant ($D = D_H + D_{OH} = 9.8$ nm$^2$/ns) give a reaction radius of $R = 0.27$ nm.[45] ($N_A$ is Avagadro's number.)

The solid line in the figure is a fit to the data using numerical integration to account for a distribution of fragment separation lengths. A Gaussian distribution centered away from the origin (i.e. $r \neq 0$) has been used to model similar dissociation reactions.[46] The best fit to the data gives an average separation length of $\langle r_{H-OH} \rangle = 0.7 \pm 0.2$ nm, regardless of whether we use a Gaussian function or a delta function distribution. Even using a Gaussian function centered at the origin, similar to the form of the electron distribution in the independent pairs model for ionization described in the previous section, gives the same value for the average separation of H and OH. A source of uncertainty in this analysis is the small contribution to the signal from the ionization products. Ignoring the contribution from ionization leads us to overestimate the recombination yield for the dissociation products because the ionization products are more likely to recombine than the dissociation products (see Fig. 3 inset). Therefore, the $r_{H-OH}$ separation length determined in our analysis is likely a lower bound.

## IV. DISCUSSION

### A. Dissociation dynamics

An important result from our measurement is the large separation of H and OH fragments following dissociation of a water molecule in the liquid ($<r_{H-OH}>$ = 0.7±0.2 nm). Dissociating H atoms escape the solvent cage and one or two additional solvent shells, suggesting that they are produced with significant initial kinetic energy and that hydrogen bonding does not drastically change the direct dissociation channel of a water molecule in the liquid relative to the gas phase. A recent molecular dynamics simulation of dissociating $H_2O$ in crystalline and amorphous ice supports this interpretation.[34,35] The simulation treats the nuclear dynamics classically using an ab initio potential energy surface for isolated water to describe the intramolecular dynamics of the dissociating molecule, a modified TIP3P model to describe its interaction with the solvent, and the TIP4P model for intermolecular interaction of bulk water molecules. By neglecting the effects of hydrogen bonding on the excited state dynamics, the simulation provides a reference point in which cooperative effects of the liquid do not significantly alter the dissociation dynamics relative to the gas phase. Our experimental result is similar to the outcome of the simulation, where H atoms travel an average distance of 0.8 nm prior to thermalization.[47] That similarity and the large separation length that we measure suggest that hydrogen bonding effects do not inhibit direct dissociation in the liquid.

On the other hand, recent high-level ab initio calculations of the first excited state of the dimer[19-21] and other small water clusters[20] suggest that hydrogen bonding has a strong influence on the shape of the potential energy surface. The calculations reveal a low barrier to dissociation along the hydrogen bond donating O-H stretch coordinate that does not exist in the isolated molecule. In one limit, such a barrier would inhibit motion along the O-H stretch, thereby dissipating excess energy from the dissociating fragments and limiting the ability of the H atom to penetrate the solvent. By inhibiting the direct dissociation channel, a barrier also potentially increases the lifetime of the excited state and reduces the quantum yield for dissociation.[20] The timescale of roughly 1 ps for hydrogen bond breaking in the electronic ground state[5,6] would give an upper limit for the excited state lifetime that is long enough for an alternate decay path to compete efficiently with direct dissociation. In the absence of a barrier, dissociation occurs on the timescale of an O-H stretch vibration (~10 fs).

The excitation energy determines the impact of such a barrier on the excited state dynamics. At low excitation energies the barrier may be insurmountable and have a profound influence on the dynamics, but at energies sufficiently high to overcome the barrier it has little effect. Indirect measurements of the dissociation yield for one-photon excitation of liquid water reflect this energy dependence. Dissociation occurs with only 45% yield for one-photon excitation at 6.7 eV,[29] indicating that a competing decay channel capable of relaxing the molecule to the ground state is important at that energy, whereas the yield increases to 70% at 8.4 eV.[28] The dissociation yield increases with energy across this range as the influence of the barrier decreases.

Our measurement of the relative yields of ionization and dissociation for two-photon excitation (see Table II) also indicate a dominant dissociation channel at 8.3 eV. However, it is important to note that the character of the excited state produced with the same excitation energy may be different for one- and two-photon transitions because of the different selection rules.[48] Even though 8.3 eV is at the center of the first absorption band in the one-photon spectrum of the liquid, two-photon excitation at this energy may also excite into the tail of higher-lying excited states. Experiments to better understand the relative two photon cross-sections are ongoing in our labs, but the similar one- and two-photon dissociation yields suggest that there is not a significant difference in the excited state for excitation at 8.3 eV. If indeed we prepare predominantly the first excited state in our experiment, the large dissociation yield implies that the barrier is relatively inefficient at that energy. A large dissociation yield is consistent with the production of H atoms with substantial kinetic energy because a barrier does not inhibit the dissociation channel. Above 8.3 eV, the dissociation yield decreases as ionization becomes increasingly important,[33] likely due to the changing character of the excited state. There are no theoretical calculations in the literature that explore the effect of hydrogen-bonding on the dissociation coordinate for higher excited states of water, although these would clearly be desirable.

It is interesting to consider the role of a barrier to dissociation in the context of the hydrogen bonding structure of liquid water.[49-52] The dissociation yield may provide a measure of the extent of hydrogen bonding in the liquid if molecules in various solvent geometries experience different barrier heights and therefore different dynamics. In this picture, the solvent environment controls the dissociation yield because the height of the barrier depends on the hydrogen bonding structure of the liquid. For instance, weak hydrogen bond donors may experience a small barrier that does little to inhibit dissociation, whereas strongly bound species in tetrahedral, ice-like configurations (with longer-lived excited states)[53] may dissociate indirectly or relax through a different mechanism. Calculations by Chipman[21] indicate that non-bonded or weakly-hydrogen bonded molecules preferentially dissociate because dissociation along a non-hydrogen bonded O-H stretch is the lowest energy path in the first excited state of the dimer. If fully-coordinated waters preferentially decay through an alternate mechanism (e.g. the quantum yield of dissociation in crystalline ice is less than unity), then the branching among channels depends on the extent of hydrogen bonding in the liquid. Temperature-dependent measurements of the dissociation yield, particularly at low excitation energies, would be informative, because raising the temperature decreases hydrogen bonding in the liquid and therefore reduces its influence on the dynamics.[54]

**B. Ionization Mechanism**

Previous studies of ionization in liquid water reveal that the mechanism depends on the excitation energy.[22-26] Direct ionization to produce $H_2O^+_{(aq)}$ and a quasi-free electron is only possible for excitation above the energy for vertical electron ejection at about 11 eV.[26] For lower excitation energies the excited molecule and its environment must reorganize in order for the system to attain a favorable geometry for electron ejection because vertical transitions to a continuum electron state are increasingly unlikely. We restrict the current discussion to excitation energies below 9.5 eV, where nuclear motion plays a central role in the ionization mechanism.[32]

The prevailing theory for ionization in this energy regime is an excited state proton-coupled electron transfer mechanism, in which electron ejection and nuclear motion along the proton transfer coordinate are simultaneous.[32] Although the exact details of this mechanism are somewhat vague, it may explain many of the experimental properties of ionization, such as the nearly constant electron ejection length $<r_0>$ below 9 eV and the exponential increase of the ionization yield with energy. The ejection length is independent of the excitation energy because that length is determined by the location of electron trap states, which does not change with energy. On the other hand, more traps are available at higher energies and the ionization yield increases accordingly. A proton-coupled electron transfer mechanism implies that ionization competes with dissociation, and that the nature of the excited state determines their relative yield.

Another proposed mechanism for low energy excitation is one in which translationally hot H atoms from dissociation collide and react with a neighboring water molecule to produce an electron and a hydronium ion.[25]

$$H + H_2O_{(l)} \rightarrow H_3O^+_{(aq)} + e^-_{(aq)} \tag{8}$$

Only sufficiently energetic H atoms react via this mechanism, which has an activation energy of about 0.7 eV.[55] An upper limit to the total available energy for the dissociating fragments is about 3.2 eV using the gas phase bond dissociation energy of 5.1 eV and the 8.3 eV excitation energy.[56] Although hydrogen bonding may alter the disposal of energy for the dissociation reaction in solution, 88% of the excess energy goes into translation of the H atom in the gas phase.[3] Thus, the reaction of H atoms with a neighboring water molecule is indeed energetically feasible if the liquid environment does not have a dramatic influence on the dissociation dynamics.

In this picture, dissociation plays an essential role in ionization, rather than being a competing process, and the efficiency of reaction (8) determines the relative branching. The increasing likelihood of H atoms having enough energy to overcome the barrier to reaction would explain the exponential increase of the ionization yield with energy. On the other hand, the location of electron trap sites in the liquid once again determines the electron ejection length, and is independent of the excitation energy.

Other reactions involving H atoms from dissociation potentially play a role as well. Two such reactions with comparable activation energies are hydrogen abstraction (0.7 eV activation energy)[57] and hydrogen exchange (0.9 eV barrier in the gas phase).[58]

$$H + H_2O_{(l)} \rightarrow H_{2\,(aq)} + OH_{(aq)} \tag{9}$$

$$H' + H_2O_{(l)} \rightarrow H'OH_{(l)} + H_{(aq)} \tag{10}$$

In the limit of ballistic H atoms, the efficiencies of reactions (8)-(10) depend on the collision energy and the impact angle of the reactants. Only the first collision or two is likely to have enough energy to overcome the reaction barrier, even for a favorable impact parameter, because collisions rapidly dissipate excess energy from the H atom.[59] The approximately tetrahedral arrangement of liquid water preferentially aligns the dissociating O-H bond toward the O atom on the nearest neighboring water molecule. Thus, the most likely first collision has an unfavorable orientation for hydrogen abstraction that limits the yield of reaction (9).[58] The transition state geometry of reaction (10), on the other hand, is a trigonal $H_3O$ molecule that is readily accessible by dissociating a water molecule in a relaxed configuration of the liquid.[58] Although speculative, it is interesting to consider that the $H_3O$ species could also be an intermediate state for ionization via reaction (8).[18]

    The excited state potential energy surface plays an important role in both of the proposed mechanisms of ionization. The essential difference between them is the role of hydrogen bonding in the excited state dynamics. In one limit, hydrogen bonding promotes excited state proton transfer, whereas in the other limit prompt dissociation similar to the gas phase produces translationally hot H atoms that react with a neighboring water molecule. Both mechanisms explain the constant ejection length and exponentially increasing ionization yield below 9 eV. The large fragment separation that we observe, indicating a direct dissociation channel, tenuously supports the hot H atom reaction mechanism of ionization, but cannot exclude the

possible role of proton-coupled electron transfer.

## V. SUMMARY

Transient absorption spectroscopy reveals the relative yields of ionization and dissociation following two-photon excitation of liquid water at 8.3 and 9.3 eV. The two decay channels occur with nearly equal probability at the higher energy, but for 8.3 eV excitation the dissociation channel dominates. The transient decay of the absorption signal at 267 nm provides information about the dissociation reaction at the lower energy, and we find that the average initial separation of H and OH fragments is 0.7±0.2 nm, about two solvent shells. This first determination of the dynamics occurring in the dissociation channel provides a new insight into the overall excited state relaxation of liquid water.

The large fragment separation that we observe implies that dissociation produces H atoms with significant initial kinetic energy. Our measurement points to a direct dissociation channel that is not significantly inhibited by a barrier on the excited state potential energy surface even though one-photon dissociation yields suggest that a barrier may play a role at lower excitation energies. The production of ballistic H atoms from dissociation tenuously supports a picture of the ionization mechanism in which H atoms react with neighboring water molecules to give the ionization products, although we cannot rule out a proton-coupled electron transfer mechanism. The hot H atom reaction mechanism is an appealing picture considering that direct dissociation of isolated water molecules produces H atoms with significant kinetic energy.

## ACKNOWLEDGEMENTS

The authors wish to thank C. D. Jonah for several helpful conversations. This work was performed at Argonne National Laboratory and was supported by the U.S. Department of Energy, Office of Science, Office of Basic Energy Sciences, under contract DE-AC02-06CH11357. S.E.B. is supported by the National Science Foundation (CHE 0311814 and 0617060).

**Table I:** Product molar extinction coefficients ($M^{-1}cm^{-1}$)

|  | 650 nm | 267 nm |
|---|---|---|
| $e^-_{aq}$ | 15 500[a] | 600[b] |
| $OH_{aq}$ | 0 | 420[b] |

[a] From Ref. 36.

[b] From Ref. 37. Although other estimates of the absorption coefficients at 267 nm vary by as much as ±15%, the difference does not have a significant impact on our result.

**Table II:** Ionization and dissociation yields for two-photon excitation

|  | Ionization yield[a] | $[OH_{aq}]/[e^-_{aq}]$[b] | Dissociation yield[c] |
|---|---|---|---|
| 8.3 eV | 12% | >8 | >84% |
| 9.3 eV | 44% | 1.7±0.9 | ~31% |

[a] From Ref. 32, extrapolated to $t = 0$ ps using the independent pairs model and the values of $<r_0>$ from the data in this work.

[b] From the acid quenching experiments in this work.

[c] Approximate dissociation yield assuming constant $[OH_{aq}]/[e^-_{aq}]$.

**FIGURE CAPTIONS**

**Figure 1.** Transient absorption following two-photon excitation at 9.3 eV. The signal at 650 nm (closed circles) is entirely due to the absorption of light by solvated electrons, whereas the absorption at 267 nm (open circles) is due to both solvated electrons and OH radicals. The line is a best fit to the electron decay using the independent pairs model. The inset shows the normalized traces.

**Figure 2.** Transient absorption in pure water (closed circles) and 2M $HClO_4$ solution (open circles) following 9.3 eV excitation. The 650 nm electron signal in the top panel decays to the baseline for the acid solution due to reaction of electrons with excess protons. The 267 nm signal in the bottom panel only decays by about 45% for the acid solution, relative to pure water, indicating that OH radicals contribute about 55% of the absorption at that wavelength.

**Figure 3.** Transient absorption at 650 nm (closed circles) and 267 nm (open circles) following two-photon excitation at 8.3 eV, with normalized traces in the inset. The line is a best fit to the electron decay using the independent pairs model.

**Figure 4.** Transient absorption in pure water (closed circles) and 2M $HClO_4$ solution (open circles) following 8.3 eV excitation. Unlike the data in Fig. 2, the 267 nm absorption does not decay in the acid solution relative to pure water because the contribution from solvated electrons is very small in this case.

**Figure 5.** Transient 267 nm absorption in pure water following two-photon excitation at 8.3 eV.

The line is a best fit to the data using the simple geminate recombination model from the text.

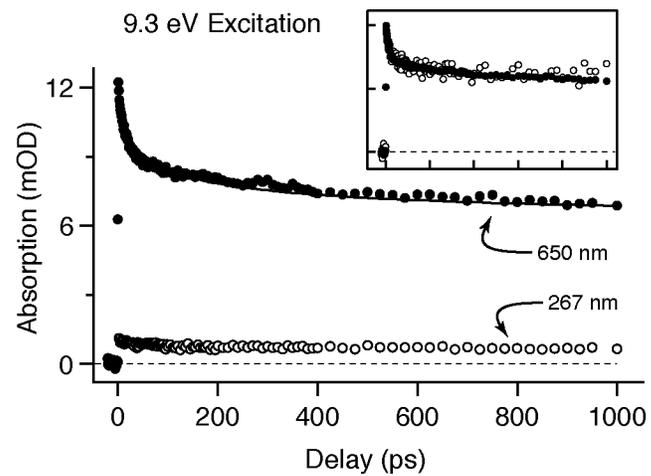

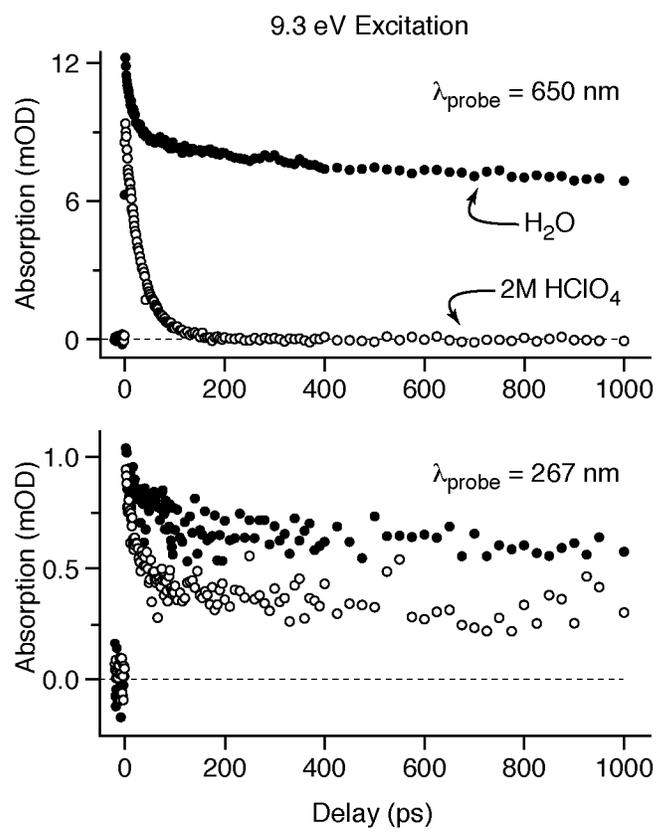

Elles, et al., Figure 2

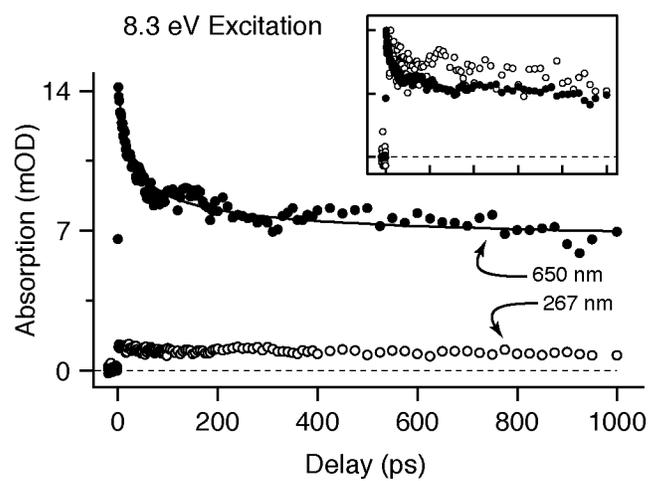

Elles, et al., Figure 3

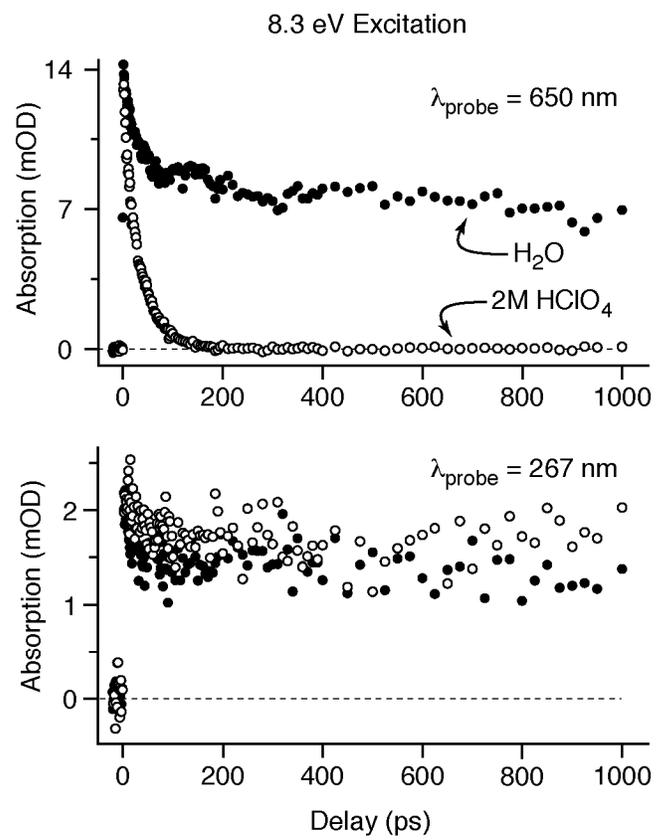

Elles, et al., Figure 4

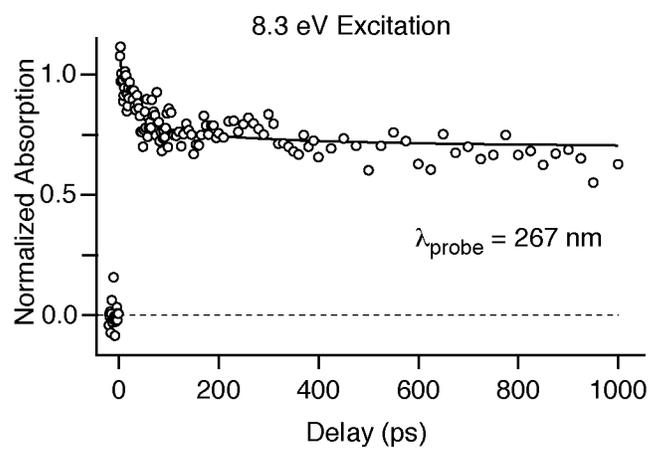

Elles, et al., Figure 5